\documentclass{PoS}

\PoS{PoS(LAT2005)128}

\title{Neutron electric dipole moment on the lattice}

\ShortTitle{Neutron electric dipole moment on the lattice}

\author{\speaker{Eigo Shintani}\thanks{shintani@het.ph.tsukuba.ac.jp} \\ 
        Graduate School of Pure and Applied Sciences, University of Tsukuba}
\author{S. Aoki\\
        Graduate School of Pure and Applied Sciences, University of Tsukuba, 
        and\\
        Riken BNL Research Center, Brookhaven National Laboratory}
\author{N. Ishizuka, Y. Kuramashi, A. Ukawa, T. Yoshi\'e\\
        Graduate School of Pure and Applied Sciences, University of Tsukuba 
        and\\
        Center for Computational Sciences, University of Tsukuba}
\author{K. Kanaya \\ 
        Graduate School of Pure and Applied Sciences, University of Tsukuba}
\author{Y. Kikukawa\\
        Department of Physics, Nagoya University, 
        , Nagoya 464-8602, Japan}
\author{M. Okawa\\
        Department of Physics, Hiroshima University}

\abstract{
We carry out a feasibility study toward a lattice QCD calculation of 
the neutron electric dipole moment (NEDM) in the presence of the $\theta$ term
using two different approaches. In the first method,
we calculate the CP-odd electromagnetic form factor $F_3$,
which becomes the NEDM in the zero momentum transfer limit.
At the first order in $\theta$, we derive a formula connecting the lattice 
three-point function to the CP-odd electromagnetic form factor.
In the second method 
we directly extract the NEDM from the energy difference 
between spin-up and spin-down neutron states in the presence of a 
constant electric field, without expanding a small but non-zero $\theta$.
We test both approaches numerically,
employing the domain-wall quark action with the RG improved gauge action
in quenched QCD at $a^{-1}\simeq 2$ GeV on a $16^3\times 32\times 16$ 
lattice, and 
further applying the second method to the clover quark action at a similar 
lattice spacing and nucleon mass.
We obtain good signals from both approaches. In particular the second 
method works well with both fermion formulations.
}

\FullConference{XXIIIrd International Symposium on Lattice Field Theory\\
		 25-30 July 2005\\
		 Trinity College, Dublin, Ireland}
\usepackage{epsfig}
\usepackage{multicol}
\begin{document}

\section{Introduction}
One of the most stringent constraints on possible violation 
of parity and time-reversal symmetry  in the strong interaction
comes from the measurement of the electric dipole moment (EDM) for neutron 
(NEDM) ${\vec d}_{n}$ and proton (PEDM) ${\vec d}_{p}$.  
The current upper bound is given by
$ |{\vec d}_n| < 6.3\times 10^{-26}\,\,e\cdot\textrm{cm}\,\textrm{(90\% C.L.)\cite{Harris}}$ and  
$|{\vec d}_p| < 5.4 \times 10^{-24}\,\,e\cdot\textrm{cm}\,\textrm{\cite{Dmitriev}}$. 
On the other hand, QCD allows a gauge invariant renormalizable CP-odd
$\theta$ term in the Lagrangian. 
Crude model estimations\cite{Crewther,QCDsum_ChPT} and the experimental bound
on $\vert \vec{d}_n\vert$ yield  a very stringent bound 
$\theta \le O(10^{-10})$.

Current model estimates give not only an order of magnitude different results 
on $\vec d_n$ but also differ even in its sign.
Clearly a first principles determination of $\vec{d}_n$ from lattice QCD 
is required to determine the value of $\theta$, if a non-zero value is 
found for NEDM in future experiments.
Indeed serious attempts toward lattice QCD calculations of NEDM have
just been started recently\cite{Shintani,Faccioli,Blum}.
In this report we present our results on NEDM calculations based on 
two different methods, one of which has already been published\cite{Shintani}.

\section{NEDM from the electromagnetic form factor}
\subsection{Formulation}
Let us consider the electromagnetic form factor of the nucleon defined by
\begin{equation}
\langle N (\vec p,s)\vert J_\mu^{\rm EM}\vert
N(\vec{p^\prime},s^\prime)\rangle 
= \bar u(\vec p,s)\left[ \frac{F_3(q^2)}{2m_N} q_\nu
\sigma_{\mu\nu}\gamma_5+\cdots
\right] u(\vec{p^\prime},s^\prime),
\end{equation}
where $J_{\mu}^{\rm EM}$ is the electromagnetic current, 
$q=p-p^\prime$ is the momentum transfer, $\vert N(\vec
p,s)\rangle$ is the on-shell nucleon state with momentum $\vec{p}$, energy
$p_0=\sqrt{m_N^2+\vec{p}^2}$ and helicity $s$.
On the right-hand side we explicitly write down the form factor $F_3$
only, which is related to NEDM as
$
\vert {\vec d}_n\vert = \displaystyle \lim_{q^2\rightarrow 0}
{F_3(q^2)}/{2m_N}={F_3(0)}/{2m_N}
$ .

The electromagnetic form factor can be extracted from 
the following three-point correlation function:
\begin{eqnarray}
G_{NJ_\mu N}^\theta (q,t,\tau) &\equiv &
{}_{\theta}\langle N(\vec p,t) J_{\mu}^{\rm EM}(\vec q,\tau)\bar 
N(\vec{p^\prime},0) \rangle_{\theta} \nonumber\\
  &=& e^{-E_{N^\theta}(t-\tau)}e^{-E_{N^\theta}^\prime t}
\sum_{s,s^\prime} u_N^{\theta}(\vec p,s)\bar u_N^{\theta}(\vec p,s)
W_{\mu}^{\theta}(q) 
u_N^{\theta}(\vec p^\prime,s^\prime)\bar u_N^{\theta}(\vec p^\prime,s^\prime)
+ \cdots, \label{eq:3-pt_in_theta} \\
W_\mu^\theta(q) &=&  \gamma_{\mu}F_1(q^2) + \frac{F_2(q^2)}{2m_N}q_{\nu}\sigma_{\mu\nu}
   + i\theta\left( \frac{F_3(q^2)}{2m_N}q_{\nu}\gamma_5\sigma_{\mu\nu}
                + F_A(q^2)(q_{\mu}q\hspace{-.2cm}/ - q^2\gamma_{\mu})\gamma_5 \right)
   + O(\theta^2)\nonumber\\
\end{eqnarray}
where $N(\vec p, t)$ and $\bar N(\vec p, t)$ are the 
interpolating fields of the nucleon at time $t$, 
and $F_1$ and $F_2$ are CP-even form factors while 
$F_3$ and $F_A$ are CP-odd.
The on-shell spinor $u^\theta_N$ satisfies
\begin{equation}
  \sum_{s} u_N^\theta (\vec{p},s)\bar u_N^\theta (\vec{p},s)
  = Z_N^{\theta} \frac{-i\gamma\cdot p + m_{N^\theta}e^{if_N(\theta)\gamma_5}}
                    {2E_{N^\theta}}
  = Z_N \frac{-i\gamma\cdot p + m_{N}}{2E_{N}} 
  + i\theta Z_N \frac{f_N^1m_N}{2E_N}\gamma_5
\label{eq:2-pt_in_theta}
\end{equation}
and for small $\theta$ we have an expansion of the forms
$m_{N^\theta} = m_N + O(\theta^2)$, $Z_N^\theta = Z_N + O(\theta^2)$ and 
$f_N(\theta) = f_N^1 \theta + O(\theta^3)$. 

In lattice simulations, the three-point function is calculated as
\begin{eqnarray}
G_{NJ_\mu N}^\theta (q,t,\tau) &=&
\int {\cal D}\,U {\cal D}\, \bar\psi {\cal D}\, \psi e^{S_{\rm QCD} + 
i\theta Q}
= G_{NJ_\mu N}(q,t,\tau) + i\theta G^Q_{NJ_\mu N}(q,t,\tau) + O(\theta^2) .
\label{eq:3-pt_lattice}
\end{eqnarray}
Comparing eq.(\ref{eq:3-pt_in_theta}) with eq.(\ref{eq:3-pt_lattice}),
we obtain two independent formulae which allow an extraction
of $F_3$.  They are  given by
\begin{eqnarray}
&&\textrm{tr }[G^Q_{NJ_4 N}(q,t,\tau)\Gamma_{4}\gamma_5]
= |Z_N|^2e^{-E_{N}(t-\tau)}e^{-E_{N}^\prime t}\nonumber\\
&&\quad\times\bigg[\frac{{\vec q}^2}{2E_Nm_N}F_3(q^2)
  + \bigg(\frac{E_N+m_{N}}{2E_N}F_1(q^2) 
   + \frac{{\vec q}^2}{4m_NE_N}F_2(q^2)\bigg)f^1_N\bigg],\label{f3_proj1}
\label{eq:F3A} \\
&&\textrm{tr }[G^Q_{NJ_4 N}(q,t,\tau)i\Gamma_{4}\gamma_5\gamma_i]
= |Z_N|^2e^{-E_{N}(t-\tau)}e^{-E_{N}^\prime t} \nonumber\\
&&\quad\times \bigg[ -\frac{E_N+m_N}{2E_Nm_N}q_iF_3(q^2)
   + \left(-\frac{q_i}{2E_N}F_1(q^2)
    -\frac{q_i(E_N+3m_N)}{4m_NE_N}F_2(q^2)\right)f^1_N\bigg]\label{f3_proj2} ,
\label{eq:F3B}
\end{eqnarray}
where $\Gamma_4=(1+\gamma_4)/2$.  The coefficient $f_N^1$, which we need 
to subtract off the mixing terms in the above formulae, 
can be extracted from the two-point correlation function as
\begin{equation}
\langle N(p,s)\bar N(p,s)Q\rangle = |Z_N|^2 e^{-E_Nt}\frac{f_N^1m_N}{2E_N}\gamma_5 , 
\end{equation}
and $F_1$ and $F_2$ can be obtained from $G_{NJ_\mu N}$.

\subsection{Numerical result}
We calculate two- and three-point functions on 730 quenched gauge configurations 
generated with the RG-improved gauge action at $\beta=2.6$, corresponding to
$a^{-1}=1.81(4)$ GeV determined from $\rho$ meson mass, on 
a $16^3\times 32$ lattice.
For the quark action, we employ the domain-wall fermion
with the fifth dimension length $N_s=16$ and
the domain-wall height $M=1.8$,
taking the quark mass of $m_f=0.03$ corresponding to $m_{PS}/m_{V}=0.63$.
The topological charge $Q$ is calculated
with the cooling method with an $O(a^2)$ improved definition.
As shown in ref.\cite{Shintani}, 730 configurations are sufficient to
extract the CP-odd phase parameter $f_N^1$ from the two-point correlation 
function, and we obtain $f_N^1 = -0.247(17)$.

\begin{figure}
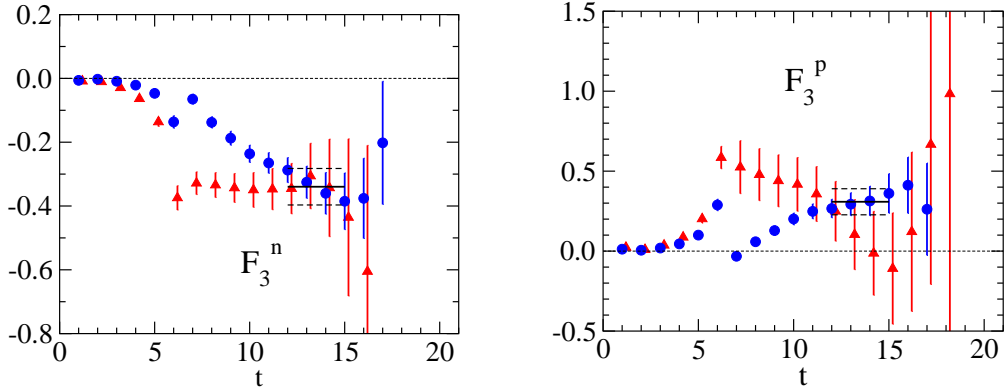

\begin{center}
\epsfig{file=Fig/f3n.leading.eps,width=.4\textwidth}
\hspace{1cm}
\epsfig{file=Fig/f3p.leading.eps,width=.4\textwidth}
\end{center}
\vspace{-.5cm}
\caption{The $F_3$ term for neutron (left) and proton (right)
as a function of $t$ with the current inserted at $\tau=6$. 
Circles represent the result with $\Gamma_4\gamma_5$ projection in 
eq.(\protect\ref{eq:F3A}), 
while triangles that with $\Gamma_4\gamma_5\gamma_j$ 
in eq.(\protect\ref{eq:F3B}), averaged over $j=1,2,3$.
Straight lines show the fitting results for the former.}
\vspace{-.2cm}
\label{fig:f_3}
\end{figure}

Subtracting the contribution composed of $F_1, F_2, f_N^1$ from the 
three-point correlation function, 
we obtain two independent estimates for $F_3$ according to
eqs.(\ref{eq:F3A}) and (\ref{eq:F3B}), which are plotted in Fig.\ref{fig:f_3}
as a function of $t$ for $\tau=6$  being fixed.
The two estimates agree at large enough $t$, demonstrating
the correctness of our formulae. 
After fitting the results from eq.(\ref{eq:F3A}) at $ 12 \le t \le 15$,
we obtain 
\begin{equation}
  F_3(q^2\simeq 0.58\mbox{ GeV}^2 )/2m_N =\left\{\begin{array}{ll}
  -0.024(5) \,\,e\cdot\textrm{fm} &\ \textrm{neutron} \\
  +0.021(6) \,\,e\cdot\textrm{fm} &\ \textrm{proton}  \\
\end{array}
\right. .
\label{eq:f3_final}
\end{equation}

\section{NEDM from the energy shift due to the electric field}
In the previous section, using the quenched lattice QCD calculation,
we successfully obtained the signal for the CP-odd form factor $F_3$. 
However, calculating NEDM still requires the limit $q^2\rightarrow 0$, 
which is not so easy in lattice QCD 
since momenta can be varied only discretely in a 
finite box and the statistical error increases as the momentum increases.
Therefore, we have investigated an alternative method proposed 
in ref.\cite{Aoki},
which does not require the $q^2\rightarrow 0$ extrapolation.

\subsection{Formulation}
A non-zero NEDM generates an energy shift of the nucleon
in the presence of a uniform and static electric field $\vec E$.  
For $\vec E = (0,0,E)$, for example, the energy difference 
between the spin-up state $N_\uparrow$ and the spin-down state $N_\downarrow$ 
takes the form
$
m_{N^{\theta}_{\uparrow}}(E)-m_{N^{\theta}_{\downarrow}}(E) = d^{\theta}_n E 
+ O(E^3) 
$ .

We include a uniform electric field by the substitution
$U_3(x)\rightarrow U_3(x)e^{qEt}$,
where the quark charge $q$ is $2/3$ for the up quark or $-1/3$ for the
down quark.
This substitution, necessary for a real electric field in Minkowski space, 
violates the periodicity in the t-direction. 
In order to minimize its effect at the boundary, 
we have to take the electric field as small as possible. 
Choosing appropriate spin components in the neutron propagator, 
we have 
\begin{equation}
\label{eq:ratio}
R(E,\theta, t) \equiv
\left[\langle N_1\bar N_1\rangle_{\theta}(E,t)\right]/
\left[\langle N_2\bar N_2\rangle_{\theta}(E,t)\right] 
= Z e^{- d^{\theta}_n E t + O(E^3)} + \cdots ,
\end{equation}
where 
\begin{equation}
\langle N_\alpha\bar N_\alpha\rangle_{\theta}(E,t) =
\langle N_\alpha (t) \bar N_\alpha (0) e^{i\theta Q}\rangle_E
\end{equation}
and $\langle {\cal O}\rangle_E$ represents the vacuum expectation value of
${\cal O}$ without the $\theta$ term but with $E$.
Since the neutron propagator calculated at
$\theta = 0$ is averaged with the weight $e^{i\theta Q}$ (``reweighting'') 
to obtain the $\theta$-average,
a good sampling of the topological charge $Q$ is crucially important
for a success of this method.

\subsection{Numerical results with the domain-wall fermion}
We take the same simulation parameters as in the previous 
section except for a heavier quark mass, $m_f=0.12$, 
corresponding to $m_N \simeq 2.2$ GeV.  The number of configurations is 
increased to 1000. 

\begin{figure}
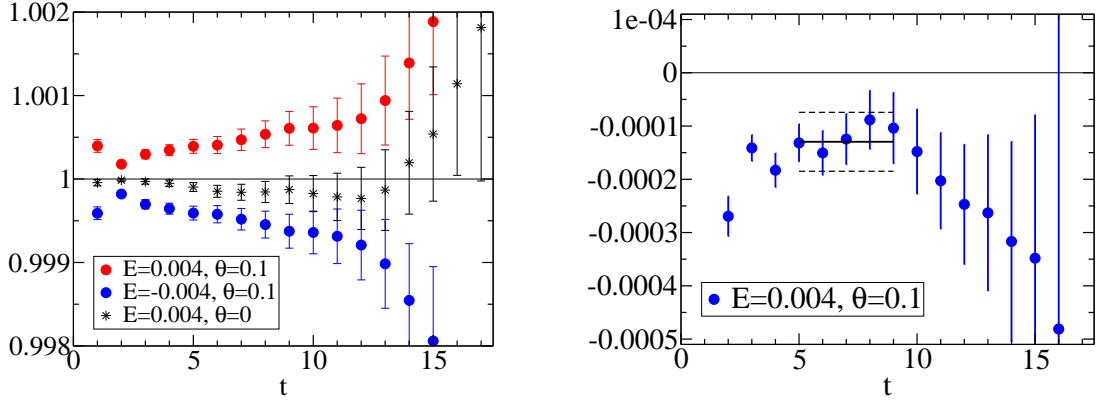

\begin{center}
\epsfig{file=Fig/NEdn.E0004.theta01.DW.eps,width=.43\textwidth}
\hspace{1cm}
\epsfig{file=Fig/NeffEdn.E0004.theta01.DW.eps,width=.45\textwidth}
\end{center}
\vspace{-.6cm}
\caption{(Left) The ratio eq.(\protect\ref{eq:ratio}) as a function of $t$ 
for $E=\pm 0.004$ and $\theta =0.1$ (circles), 
and for $E=0.004$ and $\theta = 0$ (bursts). 
(Right) An effective mass plot of the ratio.}
\label{fig:EDM_DW}
\vspace{-.2cm}
\end{figure}

On the left of Fig.\ref{fig:EDM_DW} the ratio eq.(\ref{eq:ratio}) is 
plotted as a function of $t$ for $E=\pm 0.004$ and $\theta =0.1$ (circles), 
and for $E=0.004$ and $\theta = 0$ (bursts). 
The behavior of the  ratio is consistent with the expectation 
that the exponent depends on the sign of $E$ and
that the ratio is unity for $\theta=0$.
On the right in Fig.\ref{fig:EDM_DW} we show the effective mass of the ratio, 
from which we determine the fitting range to be $5\le t\le 9$.  

\begin{figure}[b]
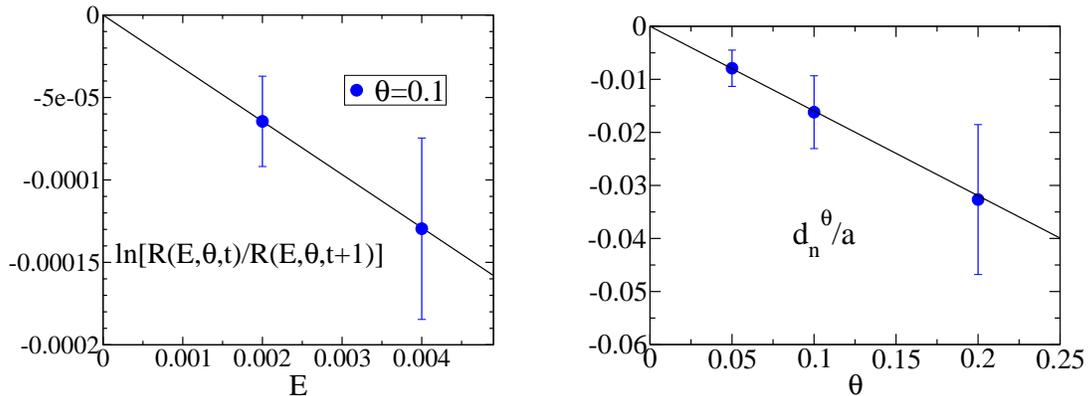

\begin{center}
\epsfig{file=Fig/NEdep.E0004.DW.eps,width=.43\textwidth}
\hspace{1cm}
\epsfig{file=Fig/Nthetadep.E0004.DW.eps,width=.44\textwidth}
\end{center}
\vspace{-.5cm}
\caption{(Left) $E$ dependence the ratio $R(t,E,\theta)$. 
(Right) $\theta$ dependence of NEDM over lattice spacing $a$.}
\label{fig:EDM_DW_E}
\vspace{-.2cm}
\end{figure}

As is seen on the left of Fig.\ref{fig:EDM_DW_E}, 
the exponent of $R(E,\theta,t)$
has a linear behavior in $E$ as expected.
Finally, in right of Fig.\ref{fig:EDM_DW_E}, $d_n^\theta$ is plotted as a 
function of $\theta$.  A slope obtained by a linear fit becomes
\begin{equation}
d_n^{\theta}/\theta = -0.0183(60)\,e\cdot\rm{fm},\quad 
d_p^{\theta}/\theta = 0.0176(72)\, e\cdot\rm{fm} .
\end{equation}
These values are smaller than those obtained with the CP-odd form factor 
at $q^2 \simeq 0.58$ GeV$^2$ ( eq.(\ref{eq:f3_final}) ).
Note, however, that the quark mass is a factor four larger in the present calculation.

\subsection{Preliminary results with the clover fermion}
\begin{figure}
\begin{center}
\epsfig{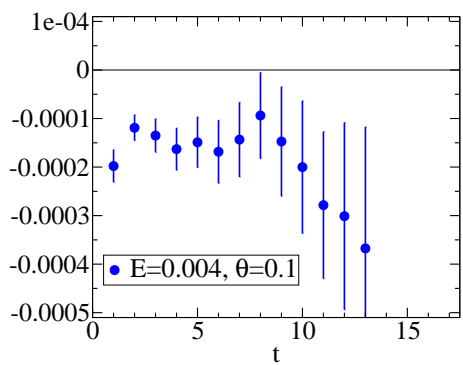}
\end{center}
\vspace{-.7cm}
\caption{An effective mass plot as Fig.2 with the clover fermion 
at $K=0.1320$,  $\theta=0.1$ and $E=0.004$.}\label{fig:EDM_clv}
\vspace{-.2cm}
\end{figure}

Since the Wilson-clover fermion action requires much less computational 
costs,  and
there exists a large number of full QCD configurations generated with 
this action,
it is important to test if the method of energy difference works with it. 
We calculate the ratio $R(E,\theta, t)$ using the gauge 
configurations generated with the same parameters, and choosing 
the hopping parameter $K= 0.1320$ which yields a nucleon mass similar 
to the domain-wall case. 
In Fig.\ref{fig:EDM_clv} we present our preliminary result for the 
effective mass.  It looks very similar to the one with the domain-wall 
action. Our results suggest that the difference of the two fermion actions
does not affect the success of this method.

\section{Conclusion and outlook}
In this report we have proposed and tested two different approaches
for the calculation of NEDM on the lattice. 
In the first method the CP-odd electromagnetic form factor is
employed, while in 
the second one the energy shift is calculated in the presence
of an electric field.
We have shown that both methods work well with the domain-wall quark action.
Furthermore we have  demonstrated that the second method also works
with the clover quark action.
We are now calculating NEDM at lighter quark masses with the clover quark
action in quenched QCD, in order to see how errors increase as the quark mass
decreases.
Our final target is the calculation of NEDM on full QCD configurations
generated with the clover quark action by the CP-PACS collaboration.

This work is supported in part by Grant-in-Aid of the Ministry of Education
(Nos. 
13640260, 
15204015, 
15540251, 
15740134, 
15740165, 
16028201, 
16540228, 
17540249, 
17540259  
).

\end{document}